# Compact Broadband Low-Loss Taper for Coupling to a Silicon Nitride Photonic Wire

PURNIMA SETHI,[1] RAKSHITHA KALLEGA,[1] ANUBHAB HALDAR,[2] AND SHANKAR KUMAR SELVARAJA[1,*]

[1]Centre for Nano Science and Engineering (CeNSE), Indian Institute of Science, Bengaluru, India.
[2]Boston University College of Engineering, Boston University, Amherst, USA.
*Corresponding author: shankarks@iisc.ac.in

**We demonstrate an ultra-compact waveguide taper in Silicon Nitride platform. The proposed taper provides a coupling-efficiency of 95% at a length of 19.5 μm in comparison to the standard linear taper of length 50 μm that connects a 10 μm wide waveguide to a 1 μm wide photonic wire. The taper has a spectral response > 75% spanning over 800 nm and resilience to fabrication variations; ±200 nm change in taper and end waveguide width varies transmission by <5%. We experimentally demonstrate taper insertion loss of <0.1 dB/transition for a taper as short as 19.5 μm, and reduces the footprint of the photonic device by 50.8% compared to the standard adiabatic taper. To the best of our knowledge, the proposed taper is the shortest waveguide taper ever reported in Silicon Nitride.**

Large-scale photonic integrated circuits (PICs) provide immense potential to meet the anticipated requirements of ultrafast computing, communication and sensing applications [1-3]. In particular, CMOS compatible materials such as Silicon, Germanium, and Silicon Nitride have evolved to address key application space such as high-speed interconnects, Mid-IR and visible-broadband photonics. Silicon Nitride (SiN) has emerged as an attractive alternative material for multiple applications due to its transparency spanning from visible to mid-IR wavelength range. These advantages has resulted in exploitation of SiN for linear and non-linear on-chip processes [4-22]; offering multitude of applications including large-scale phased arrays at both infrared and visible wavelength [12], low loss waveguiding [13], opto-mechanics [14], nonlinear optics i.e. parametric amplification [15] and broadband supercontinuum generation [16, 17] enabled by its low nonlinear loss [18], on-chip spectroscopic sensing [19], entangled photon generation [20], and hybrid photonic devices [21, 22]. However, due to its moderate refractive index contrast with Silicon Dioxide, the circuits are relatively large due to larger bend radii. Thus, realizing compact devices and circuits in SiN platform is extremely crucial.

Waveguide tapers are an elemental part of PICs and are crucial to realize coupling between devices of varying dimensions. Since the taper length primarily depends on the starting and ending waveguide width and index-contrast, the transition between a grating coupler (GC) and a single-mode photonic wire waveguide is one of the largest [23, 24]. A grating footprint of 10 μm ×10 μm is typically chosen to mode-match the grating field with a single-mode optical fiber. The grating is then coupled to a waveguide through an adiabatic taper.

Adiabatic tapers based on Silicon Nitride platform typically provide a coupling efficiency of >95% for lengths exceeding 50 μm as there is a tradeoff between the taper length and coupling efficiency due to the adiabatic transition. Consequently, the footprint of the GCs based on adiabatic tapers is limited by the length of the taper.

It is therefore tremendously essential to design compact tapers on nitride platforms to facilitate realization of dense integrated circuits. The challenge is to design a compact taper that has low-insertion loss, low-reflection and is broadband as well as robust to fabrication imperfections.

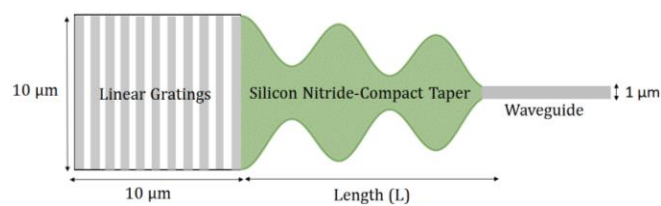

**Fig. 1.** Schematic illustration of the proposed compact taper structure.

In this paper, we experimentally demonstrate a compact high-efficiency waveguide taper between a GC and single mode waveguide in Silicon Nitride platform. The taper is defined using a quadratic sinusoidal function and is highly robust to fabrication imperfections. The taper is merely 19.5 μm long with an insertion loss as low as 0.22 dB at 1550 nm.

A schematic of the taper along with a GCs is shown in Fig. 1. Unlike an adiabatic taper, the proposed taper works on multi-mode interference along its length, resulting in maximum coupling to the fundamental waveguide mode. The proposed taper to connect a broad waveguide section to a submicron waveguide section is defined using an interpolation formula [25],

$$X = a(bz^2 + (1-b)z) + (1-a)\sin\frac{c\pi z^2}{2} \quad (1)$$

where $0 \leq a \leq 1$, $-\frac{c}{c-2} \leq b \leq \frac{c}{c-2}$, $c$ is any odd integer $\geq 3$. This formula meets the following boundary conditions: X ($z = 0$) = 0 and X ($z = 1$) = 1, where $z$ is the relative length of the taper. The final width profile, X = f ($z, a, b, c$) is a superposition of a parabolic baseline and the square of a sine.

Figure 2 shows the effect of $a$, $b$ and $c$ on the taper profile for a taper length of 19.5 μm. The Eigen Mode Expansion (EME) algorithm and Finite Difference Method (FDM) were used to optimize the taper parameters to achieve shortest taper with high-transmission between the waveguide sections. The optimal parameters were found by comprehensive numerical simulation using iterative parametric sweeps, with feedback-based parameter refinement.

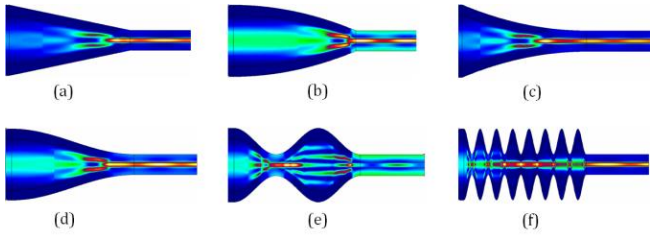

**Fig. 2.** Optical intensity profiles of the compact taper at length = 19.50 μm for (a) a = 1, b = 0 (linear taper), efficiency = 72%, (b) a = 1, b = 1, efficiency = 37%, (c) a = 1, b = -1, efficiency = 59.5%, (d) a = 0, c = 1, efficiency = 47.5%, (e) a = 0, c = 3, efficiency = 10%, and (f) a = 0, c = 15, efficiency = 54%.

Table 1 summarizes the two optimized Compact Taper (CT) configurations that yielded over 94% transmission efficiency. Figure 3(a) and (b) shows the corresponding optical intensity profiles for different optimized $b$-values and taper lengths. A maximum coupling efficiency of 95% is achieved for a taper length of 19.5 μm. Design values for $a$ and $c$ are fixed at 0.45 and 5 respectively. Subsequently, $b$ is used for fine-tuning the optimal response. Figure 3(c) illustrates the evolution of modes along the length (L) of the compact taper (19.5 μm). Figure 4(a) depicts the spectral response of the compact taper based linear GCs for the broadband 1000 nm wavelength range and (b) shows in the C+L band for all two configurations. The proposed taper has a broadband operation with > 75% transmission for a wavelength range of 800 nm covering O, C, L-band and beyond.

Additionally, the effect of dimensional variation on the transmission performance was also considered to take fabrication tolerances into account. Figure 4(c) shows the effect of end waveguide variation on the transmission. The taper transmission is unaffected for waveguide width variation of >200 nm. In practice, one can expect a width variation of ±25 nm, which results in a transmission degradation of merely < 1%. Figure 4(d) shows the effect of the total taper width variation on the coupling efficiency. As is evident, the tapers are very resilient (> 88% efficiency for ± 400 nm shift in optimized taper width).

**Table 1.** Variation in Compact Taper (CT) transmission for different $b$ values at 1550 nm. Start and end waveguide width is 10 μm and 1 μm respectively.

| Configuration | a | b | c | Length (L) μm | Transmission % | Loss dB |
|---|---|---|---|---|---|---|
| CT1 | 0.45 | 1.10 | 5 | 19.50 | 95.09 | 0.22 |
| CT2 | 0.45 | 1.15 | 5 | 19.50 | 94.90 | 0.23 |

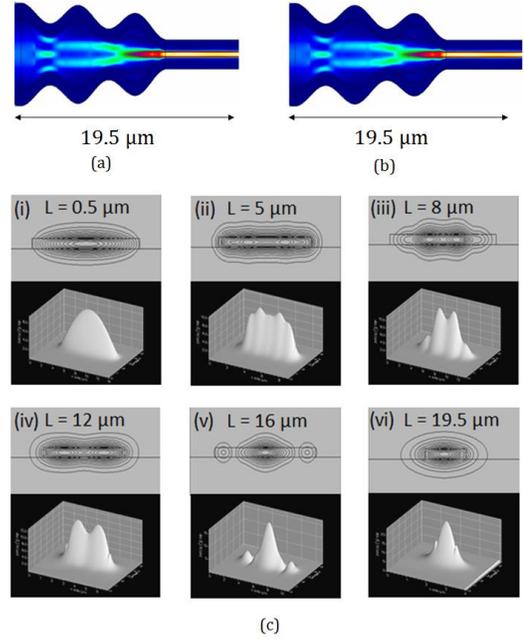

**Fig. 3.** Optical intensity profiles for the compact taper at 1550 nm TE polarization for different $b$ values and taper lengths (a) CT 1, (b) CT 2 (Table 1), and (c) 2-D Contour plot and 3-D Mode Profile along the length (L) of a compact taper CT1.

To compare the proposed taper performance with the standard adiabatic taper test structures were designed and fabricated. We have fabricated three types of structures to characterize the performance of the tapers; (a) a 10 μm wide waveguide with grating couplers, (b) grating couplers with adiabatic taper (length = 50 μm) and (c) compact taper (length = 19.5 μm). The structures were designed with an input GC along with the tapers coupling into a 1 μm wire waveguide (fiber to waveguide coupling) and tapering-out to an identical output coupler configuration (waveguide to fiber coupling). The 10 μm wide patch waveguides were designed to extract the insertion loss of the GC alone. All the GCs were designed for TE-polarized 1550 nm with a grating period of 550 nm and 50% fill-factor. The test structures were fabricated using electron-beam lithography and Inductively Coupled Plasma-Reactive Ion Etching. Patterning was done on a 400 nm thick Plasma-Enhanced Chemical Vapor Deposition (PECVD) Silicon Nitride deposited on a Bragg layer to improve the coupling efficiency. The Bragg layer was formed by alternate layers of Silicon dioxide (270 nm) and amorphous Silicon (110nm) deposited using PECVD. A 2 μm thick PECVD oxide was deposited as top cladding. Fig. 5 shows an SEM image of the grating and the CT before oxide deposition.

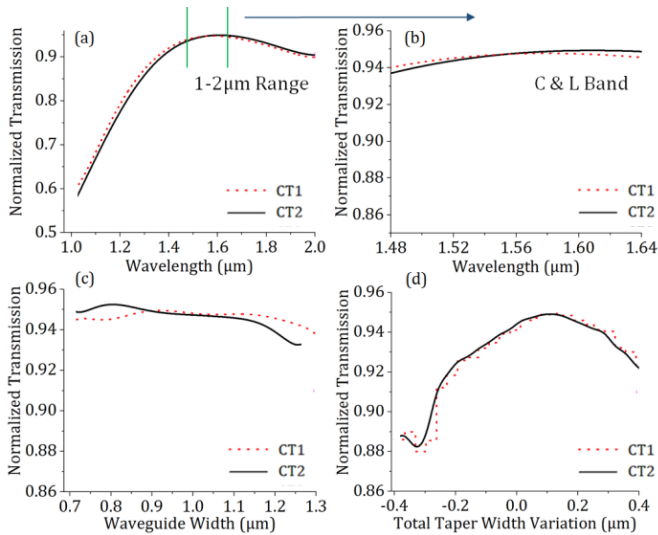

**Fig. 4.** Spectral response and tolerance of the proposed compact taper (Table 1), (a) spectral response of the compact taper in the broadband 1000 nm range (b) C & L-band (1480 nm – 1640 nm), (c) effect of end waveguide width variation on the transmission, (d) effect of compact taper width variation (different *b* values) on the transmission efficiency.

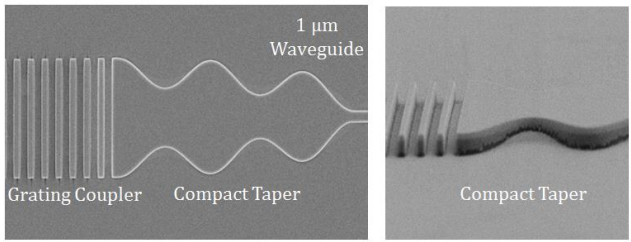

**Fig. 5.** SEM image of the compact taper along with a linear GC for 1550 nm TE polarization.

The fabricated devices were characterized using a broadband SLED (Super-luminescent Light Emitting Diode) at 1550 nm with a 3 dB bandwidth of 90 nm and an OSA (Optical Spectrum Analyzer). The polarization of the light from the SLED source is controlled using polarization wheels before the input GC. Figure 6 shows the spectral response of the fabricated devices. Figure 6(a-i) compares the performance of the compact taper with the long adiabatic taper while 6(a-ii) compares the efficiency of the two proposed compact taper configurations. Table 2 summarizes the taper performance.

As is evident from Fig. 6(a-i), the characterization results confirm that the proposed CT1 yields same coupling efficiency as the adiabatic taper. The insertion loss per coupler is 4.18 dB and 4.14 dB for GC with compact taper and adiabatic taper respectively. Fig. 6(b-i) and Fig. 6(b-ii) show the response of the proposed taper and adiabatic taper by subtracting the patch response. Using the patch waveguides, after subtracting the coupler loss, we observe insertion loss of <0.1 dB and <0.14 dB per taper for a compact taper and adiabatic taper respectively. The performance of the proposed taper is marginally better that the adiabatic with 61% reduction in length. Figure 6(a-ii) shows the compact taper's response with different design parameter *b* (Table 1). Measurement results show that a taper width variation of 30 nm would only vary the coupling efficiency by <0.2 dB, which illustrates the robustness of the proposed taper.

**Table 2. Experimental Analysis of the Performance-Metrics of the two Configurations.**

| Fiber to Waveguide Coupling | Waveguide to Fiber Coupling | Taper length [μm] | 3 dB Bandwidth (nm) | Taper Loss (dB) |
|---|---|---|---|---|
| Compact Taper (CT1) | Compact Taper (CT1) | 19.5 | ~65 | ~0.1 |
| Compact Taper (CT2) | Compact Taper (CT2) | 19.5 | ~68 | ~0.4 |
| Long Taper | Long Taper | 50 | ~62 | ~0.13 |

In conclusion, waveguide tapers are an essential part of a PIC and are necessary to realize coupling between devices of varying dimensions. We have designed and demonstrated the shortest tapered spot-size converters to couple light to a single mode waveguide from a 10 μm wide waveguide. The device shows a 50.8% reduction in the footprint compared to an adiabatic taper (600 μm²) and has a footprint of 295 μm², and is highly fabrication tolerant as well.

**Funding:** Defense Research and Development Organization, GoI-RIP/ER/1401134/M/01/1864/D(R&D)/1 581 and Science and Engineering Research Board, GoI.

**Acknowledgments:** The authors would like to thank Defense Research and Development Organization, Government of India and Office of the Principal Scientific Advisor to Government, NNFC and MNCF at the Indian Institute of Science-Bengaluru for their assistance. We would also like to thank Abhai Kumar and Siddharth Nambiar for the design of the gratings.

**References**

1. D. Perez, I. Gasulla, L. Crudgington, D. J. Thomson, A. Z. Khokhar, K. Li, W. Cao, G. Z. Mashanovich, and J. Capmany, Nat. Commun. **8**, 1 (2017).
2. D. J. Moss, R. Morandotti, A. L. Gaeta, and M. Lipson, Nat. Photon. **7**, 597 (2013).
3. D. Dai, J. Bauters, and J. E. Bowers, Light: Sci. Appl. **1**, 1(2012).
4. A. Rahim, E. Ryckeboer, A. Z. Subramanian, S. Clemmen, B. Kuyken, A. Dhakal, A. Raza, A. Hermans, M. Muneeb, S. Dhoore, Y. Li, U. Dave, P. Bienstman, N. L. Thomas, G. Roelkens, D. V. Thourhout, P. Helin, S. Severi, X. Rottenberg, R. Baets, J. of Lightwave Technol. **35**, 639 (2017).
5. L. Chang, M. H. P. Pfeiffer, N. Volet, M. Zervas, J. D. Peters, C. L. Manganelli, E. J. Stanton, Y. Li, T. J. Kippenberg, and J. E. Bowers, Opt. Lett. **42**, 803 (2017).
6. Z. Shao, Y. Chen, H. Chen, Y. Zhang, F. Zhang, J. Jian, Z. Fan, L. Liu, C. Yang, L. Zhou, and S. Yu, Opt. Express **24**, 1865 (2016).
7. C. Lacava, S. Stankovic, A. Z. Khokhar, T. D. Bucio, F. Y. Gardes, G. T. Reed, D. J. Richardson, and P. Petropoulos, Scientific Reports **7**, 22 (2017).
8. S. Kuanping, S. Pathak, G. Liu, S. Feng, S. Li, W. Lai, and S. J. B. Yoo, Opt. Express **25**, 10474 (2015).
9. M. Karpov, H. Guo, A. Kordts, V. Brasch, M. H Pfeiffer, M. Zervas, M. Geiselmann, and T. J. Kippenberg, Phys. Rev. Lett. **116**, 103902 (2016).
10. S. Kim, K. Han, C. Wang, J. A. J-Villegas, X. Xue, C. Bao, Y. Xuan, D. E. Leaird, A. M. Weiner, and M. Qi, Nat. Commun. **8**, 372 (2017).
11. S. R. García, F. Merget, F. Zhong, H. Finkelstein, and J. Witzens, Opt. Express **21**, 14036 (2013).
12. C. Poulton, M. Byrd, M. Raval, Z. Su, N. Li, E. Timurdogan, D. Coolbaugh, D. Vermeulen, and M. Watts, Opt. Lett. **42**, 21 (2017).
13. W. Sacher, Y. Huang, G. Lo, and J. Poon, J. Lightwave Technol. **33**, 901 (2015)

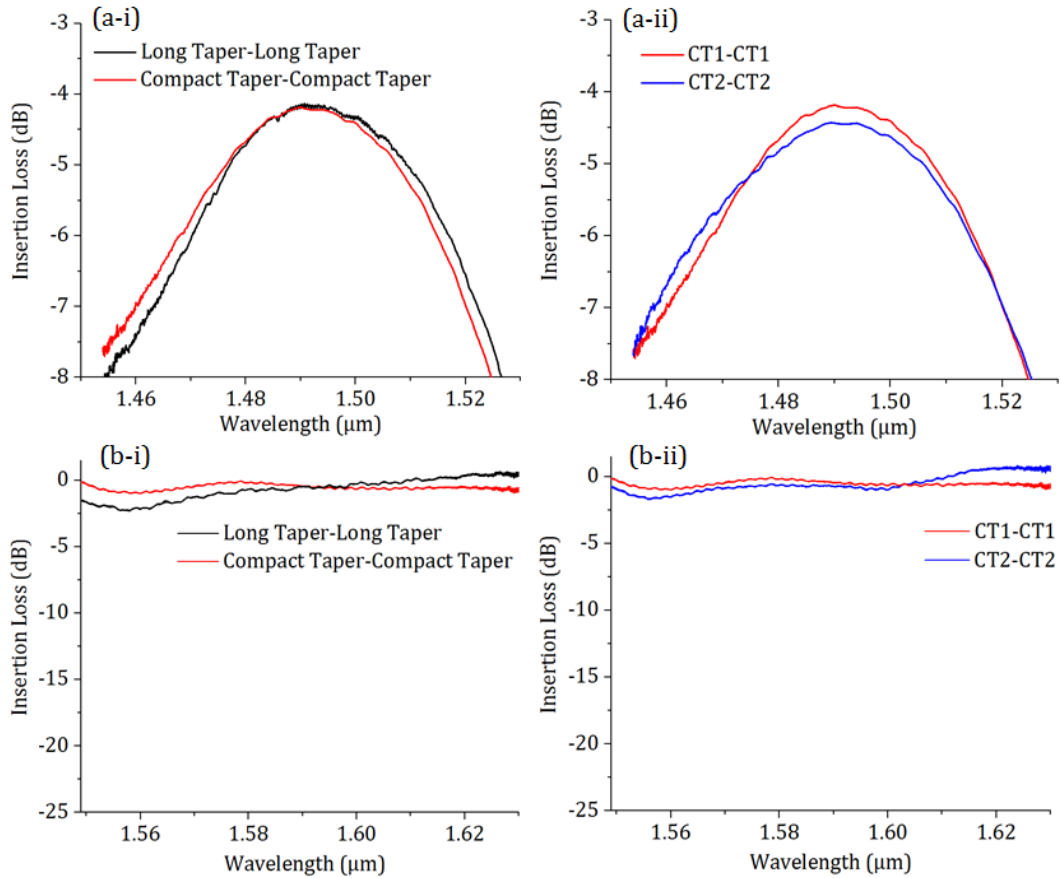

Fig. 6. (a-i) Coupling efficiency of the adiabatic (long) and proposed (compact) taper configurations of GC (Table 2), (a-ii) Effect of design parameters on the coupling efficiency of the proposed compact taper for two variations as shown in Table 1, (b-i) Insertion loss of the taper alone by neglecting the loss of the GC though a patch structure (b-ii) Insertion loss of the two proposed taper configurations (Table 1).

14. M. Eichenfield, R. Camacho, J. Chan, K. J. Vahala, and O. Painter, Nature **459**, 550 (2009).
15. J. S. Levy, A. Gondarenko, M. A. Foster, A. C. Turner-Foster, A. L. Gaeta, and M. Lipson, Nat. Photonics **4**, 37 (2010).
16. R. Halir, Y. Okawachi, J. S. Levy, M. A. Foster, M. Lipson, and A. L. Gaeta, Opt. Lett. **37**, 1685 (2012). S
17. T. Wang, D. K. Ng, S-K. Ng, Y-T Toh, A. K. L. Chee, G. FR Chen, Q. Wang, and D. TH Tan, Laser & Photonics Reviews **9**, 498 (2015).
18. T. J. Kippenberg, R. Holzwarth, and A. Diddams, Science **332**, 555 (2011).
19. A. Subramanian, E. Ryckeboer, A. Dhakal, F. Peyskens, A. Malik, B. Kuyken, H. Zhao, S. Pathak, A. Ruocco, A. De Groote, P. Wuytens, D. Martens, F. Leo, W. Xie, U. Dave, M. Muneeb, P. Van Dorpe, J. Van Campenhout, W. Bogaerts, P. Bienstman, N. Le Thomas, D. Van Thourhout, Z. Hens, G. Roelkens, and R. Baets, Photon. Res. **3**, B47 (2015).
20. C. Xiong, X. Zhang, A. Mahendra, J. He, D-Y. Choi, C. J. Chae, D. Marpaung, A. Leinse, R. G. Heideman, M. Hoekman, and C.G.H. Roeloffzen, Optica **2**, 724 (2015).
21. J. T. Bovington, M. J. R. Heck, and J. E. Bowers, Opt. Lett. **39**, 6017 (2014).
22. S. Zhu, G. Q. Lo, and D. L. Kwong, Opt. Express **21**, 23376 (2013).
23. Y. Chen, T. D. Bucio, A. Z. Khokhar, M. Banakar, K. Grabska, F. Y. Gardes, R. Halir, Í. M-Fernández, P. Cheben, and J-J He, Opt. Lett. **42**, 3566 (2017).
24. A. Z. Subramanian, S. Selvaraja, P. Verheyen, A. Dhakal, K. Komorowska, and R. Baets, **24**, 1700 (2012).
25. P. Sethi, A. Haldar, and S. K. Selvaraja, Opt. Express **25**, 10196 (2017).